\documentstyle[12pt]{article}

\input{epsf.sty}
\def\framegraphics{\def\ifframe{\iftrue}}
\def\dontframegraphics{\def\ifframe{\iffalse}}
\def\drawgraphics{\def\ifdraw{\iftrue}}
\def\dontdrawgraphics{\def\ifdraw{\iffalse}}

\dontframegraphics
\drawgraphics

\newcommand{\graphics}[6]{
\def\epsfsize##1##2{#6##1}
\begin{picture}(#2,#3)
  \ifframe
    \put(0,0){\framebox(#2,#3){}}
  \fi
  \ifdraw
    \put(0,#3){\begin{picture}(0,0)
                 \put(#4,#5){\epsfbox{#1}}
               \end{picture}}
  \fi
\end{picture}}

\newcommand{\be}{\begin{equation}}
\newcommand{\ee}{\end{equation}}
\newcommand{\bear}{\begin{eqnarray}}
\newcommand{\ear}{\end{eqnarray}}

\newcommand{\sla}{\slash\hspace {-0.5em}}

\begin{document}

IFUSP/P - 1276  \hspace*{6cm} August 1997

\vspace{2cm}
\centerline{\large Heavy Quarks in Polarized Structure Functions}
\vspace{.4cm}
\centerline{F.M. Steffens\footnote{Email address:fsteffen@axpfep1.if.usp.br}}
%\maketitle
\vspace{1cm}
\centerline{Instituto de F\'{\i}sica}
\centerline{Universidade de S\~ao Paulo}
\centerline{C.P. 66318}
\centerline{05389-970, S\~ao Paulo, Brazil}
\vspace{.3cm}
\begin{abstract}

Quark mass effects are included in the calculation of polarized structure 
functions. In particular, the validity of fixed 
order perturbation theory and of massless evolution is studied 
in the framework of heavy quarks structure functions. 
The polarized version of the ACOT and MRRS interpolating schemes
for the evolution of massive quarks distributions is also 
developed and studied. The different behaviours of the various
approaches in $x$ and $Q^2$ are shown.

\end{abstract}

\section{Introduction}

There has been a great interest in ways to improve 
our knowledge of the heavy quark content of the structure functions,
both theoretically \cite{aivazis94,olness97,buza96b,buza96,smith97,mrrs96} 
and experimentally \cite{expt}. It is an important
research subject for two main reasons: 1) The gluon distribution
in the proton has been extensively studied through heavy 
quark lepton production; 2) There are theoretical uncertainties
on how to determine heavy quark distributions in the threshold 
region and how to resum the large logs containing the quark masses.

So far, unpolarized electroproduction has been the major focus
of attention. The reason behind this activity is the large 
charm component of $F_2$ measured at HERA \cite{expt}. 
However, future experiments dedicated to measure $\Delta g$ \cite{compass96},
the polarized gluon distribution of the proton, will also 
probe the polarized charm distribution through open charm 
production. Hence the importance of a proper treatment of massive
quarks in polarized Deep Inelastic Scattering (DIS). In earlier 
attempts to estimate the charm component of $g_{1p}$, attention
was mainly restricted to its anomalous contribution 
\cite{ellis88,bass90,me96} and to the magnitude of $\Delta c$,
the charm singlet axial charge \cite{manohar88,altarelli90,we97}. 
In particular,
in Ref. \cite{me96} the $x$ dependence of the anomalous gluon
contribution to heavy quarks was calculated. One of the conclusions
of that work was that the 
inclusion of charm was quite relevant to the magnitude of 
$\Delta g$. 

In any case, a detailed study of the $x$ dependence and of the evolution
of polarized heavy quarks is still missing. There has been an 
early work of Buza et al. \cite{buza96} where NLO corrections
to the Photon Gluon Fussion (PGF) cross section is calculated
and their $x$ dependence studied. However, as it will be clear soon, 
it is desirable to expresses the heavy quark component of 
$g_{1p}$ in terms of heavy quark distributions. The main reason
for this are the large logarithms which necessarily appear in the 
partonic cross sections. In this line, it is the purpose of this
work to investigate how and when massive quarks can be treated 
as partons in polarized DIS.

In massless $QCD$, the polarized structure function $g_{1p} (x,Q^2)$ of
the proton can be written as:

\bear
g_{1p} (x, Q^2) &=& \int_x^1 \frac{dz}{z} \{ C_q^{NS} (z, Q^2, \mu^2)
\Delta q^{NS} (x/z, \mu^2) \nonumber \\*
   &+& \frac{1}{2n_f} \sum_{k=1}^{n_f} e_k^2 
   [(C_q^S (z, Q^2, \mu^2) \Delta\Sigma (x/z, \mu^2) \nonumber \\* 
   & & + C_g (z, Q^2, \mu^2) \Delta g(x/z, \mu^2)] \},
\label{11}
\ear
where the quark nonsinglet and singlet coefficient functions are

\be
C_q^{NS,S} (z, Q^2/\mu^2) = \delta (z - 1) + \frac{\alpha_s(\mu^2)}{4\pi}
 C_q^{(1), NS,S} (z, Q^2/\mu^2) + ...,
\label{12}
\ee
while the gluon coefficient function is

\be
C_g (z, Q^2/\mu^2) = \frac{\alpha_s(\mu^2)}{4\pi} C_g^{(1)} (z, Q^2/\mu^2) + ... ,
\label{13}
\ee
and the sum runs over the light flavours.
The singlet quark distribution is 

\be
\Delta\Sigma (z, \mu^2) = \sum_{i=1}^{n_f} \Delta q_i (z, \mu^2),
\label{14}
\ee
and the nonsinglet quark distribution is

\be
\Delta q_{NS} = \frac{1}{2}\sum_{i=1}^{n_f}\left( e_i^2 - \frac{1}{n_f} \sum_{k=1}^{n_f} 
e_k^2 \right) \Delta q_i (z, \mu^2).
\label{15}
\ee
We set the renormalization scale to be equal to the factorization
scale, $\mu^2$.
The $Q^2/\mu^2$ dependence appearing in Eqs. (\ref{12}) and 
(\ref{13}) is in the form of a logarithm.
Given the $x$ dependence of the axial charges in 
Eqs. (\ref{14}) and (\ref{15}) at some scale $\mu^2$, we can calculate
the structure function at any other scale $Q^2$ with the 
help of the perturbatively calculated coefficient functions
and splitting functions.
Basically, there are two ways to perform this evolution. First, 
we could fix $\mu^2$ and do the evolution via the logarithms
appearing in the coefficient functions, a procedure which is
called Fixed Order Perturbation Theory (FOPT) \cite{ziljstra94,gluck94}. 
This method becomes problematic when $Q^2 >> \mu^2$ because large
logarithms appear. Indeed, $C_g$ in Eq. (\ref{13}) is proportional 
to $\alpha_s (\mu^2)ln((1-z)Q^2/z\mu^2)$ and the LO approximation 
will be reliable as long as this term is much smaller than unity.
On the other hand, in FOPT
the structure functions are determined to any order once the
relevant partonic cross section is calculated to that order. 
It is not necessary to know all the splitting functions either to
evolve the parton distributions.
Otherwise, we could set $\mu^2$ to the natural scale $Q^2$ and
do the evolution via the DGLAP equations for 
$\Delta q^{NS} (z, Q^2)$, $\Delta\Sigma (z, Q^2)$, and $\Delta g (z, Q^2)$. 
In other words, the logarithms
in $Q^2$ would be resummed. This procedure would be the most adequate
as in the case of FOPT we have always the problem of not knowing
up to which $Q^2$ the evolution through the logarithms can be trusted.

The situation is more involved when we deal with massive quarks.
To begin with, we have an extra scale in 
the coefficient functions, the heavy quark mass $m_h$. For $Q^2 
>> m_h^2$, the mass sole effect would be to replace $\mu^2$ by $m_h^2$
in the logarithms of the coefficient functions, which means that the
quark mass is being
used as the infrared regulator. The corresponding large
logarithms can then be resummed in a similar fashion to the massless
quarks. Indeed, this has been done to all orders of perturbation 
theory by Buza et al. \cite{buza96b} for the unpolarized structure 
functions.
When $Q^2 \sim m_h^2$, this resummation can not
be done. 
It then seems natural to use FOPT to calculate the structure
function in the threshold region, using the full mass dependent
form for the coefficient function.
However, it is desirable to have a scheme that interpolates 
the two regions. The recently proposed methods of Aivazis et al. 
(ACOT) \cite{aivazis94} and of Martin et al. (MRRS) \cite{mrrs96} 
do this interpolation.
In the MRRS scheme, the relevant splitting functions are 
modified in the threshold region in order to incorporate mass 
effects in the evolution of the massive quark distributions.
It means that a heavy quark distribution is introduced at 
threshold which is evolved with the help of the relevant mass 
modified splitting functions. 
Moreover, the gluon coefficient function is modified
to avoid double counting. When $Q^2 >> m_h^2$, their expressions
are reduced to the massless convolution formulas. 
In the ACOT formalism, the splitting
functions preserve their massless form
but the coefficient functions retain their full dependence in 
$m_h^2$ to all orders of perturbation theory. Incidentally, we will
see in the present work that the ACOT scheme is natural in 
polarized DIS.

It is the aim of this paper to study the evolution
of the heavy quark component of $g_{1p}$ using three different
approaches. In section 2, the basics of the FOPT and of the massless 
evolution of $g_{1p}^h (x, Q^2)$ are spelled out.
In section 3 we calculate the modifications that 
the quark masses induce in the polarized splitting and coefficient 
functions. We then implement the interpolation schemes proposed in 
\cite{aivazis94,mrrs96}.  
The different methods for calculating $g_{1p}^h (x, Q^2)$
are compared in section 4, and their
validity is discussed in section 5.

\section{$g_{1p}^h (x, Q^2)$ in FOPT and in Massless Evolution}

In the absence of intrinsic heavy quarks in the proton, the 
dominant process for heavy quark production is PGF. This 
partonic cross section starts at order $\alpha_s$, and once
it is known, one can in principle calculate the corresponding
structure function at any scale. Hence,
if at the scale $\mu^2$ the heavy quark distribution is zero, 
then the heavy quark component of the polarized structure 
function $g_{1p} (x, Q^2)$ to order $\alpha_s$ is given by:

\be
g_{1 p}^h (x, Q^2) = e_h^2 \frac{\alpha_s (\mu^2)}{4\pi} 
  \int_{ax}^1 \frac{dz}{z}
  H_g^{(1)} (Q^2, m_h^2, z) \Delta g (\mu^2, x/z),
\label{21}
\ee
where the PGF cross section for a massive particle, $H_g^{(1)}$, is
\cite{watson82,grv91}:

\be
H_g^{(1)} (Q^2, m_h^2, z) = T_f[4(2z - 1)ln\left(\frac{1 + sq}{1 - sq}
\right) + 4(3 - 4z)sq],
\label{22}
\ee
with $T_f = 1/2$ and 

\be
sq = \sqrt{1 - \frac{4m_h^2}{(1 - z)Q^2}}.
\label{23}
\ee
The factor $a = 1 + 4m_h^2 /Q^2$ in (\ref{21}) ensures that the photon
carries enough energy for the resolution of a heavy quark pair.

The questions to be answered are: given the scale $\mu^2$, how good
is Eq. (\ref{21}) to describe $g_{1p}^h (x, Q^2)$? Up to which 
$x$ and $Q^2$ is Eq. (\ref{21}) valid?
In the absence of data, the answer to these questions should be in
the comparison with different theoretical approaches. 
The main constraint to Eq. (\ref{21}) is $\alpha_s(\mu^2) ln(Q^2/m_h^2)
<< 1$, a condition that is fulfilled in the threshold region ($Q^2 \sim m_h^2$).
However, when $Q^2 >> m_h^2$ Eq. (\ref{22}) is reduced to:

\be
H_g^{(1)} (Q^2, m_h^2, z) = T_f[4(2z - 1)ln\left(\frac{1 -z}{z}
\frac{Q^2}{m_h^2}\right) + 4(3 - 4z)],
\label{24}
\ee
and the large logarithms, mentioned in the introduction, appear. 
In that case, FOPT is not reliable any more. In an early study 
\cite{ziljstra94} it was
argued that $Q^2/m_h^2 < 100$ would be a good approximation to FOPT. As 
we will see later, this is true only in the $x$ region studied
in that work ($x > 0.01$). The discrepancy between FOPT and DGLAP evolution
in fact grows as $x$ becomes smaller, becoming as different as 100\% already
at $x=10^{-4}$.

Of course, one could circumvent the problem with the large logarithms
assuming massless 
evolution for heavy quarks over the entire $Q^2$ region. 
This approximation has been used in the literature for the 
unpolarized heavy quark distributions \cite{mrs94,cteq95}. In such case,
a heavy quark distribution function is introduced at a given 
scale, say $m_h^2$. Its value 
at any other scale is then predicted by the usual DGLAP equations.
In this approach, the heavy quarks are resummed as massless particles
and $g_{1 p}^h$ would be given by:

\bear
g_{1 p}^h (x, Q^2) &=& e_h^2 \int_{x}^1 \frac{dz}{z}
[ C_q (z, \alpha_s (Q^2))\Delta h(x/z, Q^2) \nonumber \\*
     &+& \frac{\alpha_s (Q^2)}{4\pi} C_g^{(1)} (z) \Delta g (x/z, Q^2)],
\label{25}
\ear
with $\Delta h(z, m_h^2) = 0$ and
$C_g^{(1)}$ given by Eq. (\ref{24}) taking $Q^2 = m_h^2$. The 
massless quark coefficient at $O(\alpha_s)$ is

\be
C_q (z, \alpha_(Q^2)) = \delta (z - 1) + \frac{\alpha_s (Q^2)}{4\pi}
    C_q^{(1)} (z),
\label{26}
\ee
with \cite{kodaira79,ziljstra94}

\bear
C_q^{(1)} (z) &=& C_F\left[4\left(\frac{ln(1-z)}{1-z}\right)_+ 
 -3\left(\frac{1}{1-z}\right)_+ - 2(1+z)ln(1-z) \right. \nonumber \\*
  &-& \left. \frac{1+z^2}{1-z}ln z + 4 + 2 z + \delta(1-z)
  (-4\zeta (2) - 9)\right].
\label{27}
\ear

Although massless evolution would give the correct answer at
$Q^2 >> m_h^2$, it does not solve the problem of evolution
of heavy quark distributions. To treat this problem, we apply the
interpolating method of Refs. \cite{aivazis94,mrrs96} 
to the polarized case.

\section{$g_{1p}^h$ in an Interpolating Scheme}

As massless evolution for massive particles would be clearly
wrong when $Q^2 \sim m_h^2$, and FOPT would present problems at
very large values of $Q^2$, it is desirable to have an scheme 
that interpolates between both. This procedure should 
reproduce Eq. (\ref{21}) at the threshold and Eq. (\ref{25}) 
when $Q^2 >> m_h^2$. Despite the fact that the ACOT scheme \cite{aivazis94}
does this interpolation, in a 
recent work Martin et al. \cite{mrrs96} developed an alternative
scheme where such interpolation is also obtained. The main ingredients
of their work are: 1) A heavy quark distribution is introduced at threshold,
and its evolution is according to DGLAP type equations, with 
mass modified splitting functions; 2) The PGF cross section is
modified to account for  similar contributions coming from the
evolution of the heavy quark distribution. This treatment ensures
the absence of double counting, in the same spirit of the ACOT work.

As in our work we are interested in the polarized heavy quark 
component of the structure functions, we have to consider 
the mass modifications in the polarized splitting 
and coefficient functions in order to implement the MRRS
scheme. We start computing 
the polarized gluon splitting function
$\Delta P_{hg}$ in LO.
The decay probability of the gluon into a polarized quark
pair, where the quark carries a fraction $z$ of the gluon
momentum is given by:

\be
dP_{hg}(z)dz = \frac{\alpha_s}{2\pi}\Delta P_{hg}dz dln(p_\perp^2
  + m_h^2)
\label{221}
\ee
with the polarized splitting function being defined as:

\be
\Delta P_{hg} (z) = \frac{z(1-z)}{2} \frac{A}{p_\perp^2 + m_h^2}.
\label{222}
\ee
As usual, the four vectors will be defined in the Infinite Momentum
Frame (IMF).
The amplitude $A$ for the decay, where the parent gluon is taken 
to be always in a state of positive helicity, is:

\be
A_{\pm +} = Tr[(\sla k_h + m_h)\gamma_\mu(\sla k_{\overline h}
- m_h)\gamma_\nu \frac{1\pm \gamma_5}{2}]\epsilon_+^{* \mu} 
\epsilon_+^\nu.
\label{223}
\ee
The gluon polarization vector is:

\be
\epsilon_\pm = (0, 0, \sqrt{1/2}, \pm i \sqrt{1/2}),
\label{224}
\ee
and the quark vectors are

\be
k_h = (zp + \frac{p_\perp^2 + m_h^2}{2zp}, zp, \vec p_\perp),
\label{225}
\ee
with a similar expression for $k_{\overline h}$, just $z$ replaced
by $1-z$. We also use the convention $\varepsilon^{0123} = +1$.
After a straightforward calculation, we get:

\bear
P_{h+g+} &=& T_f \frac{z(1-z)}{2} \frac{A_{++}}{(p_\perp^2 + m_h^2)^2}
\nonumber \\*
         &=& T_f \left(z^2 + \frac{m_h^2}{p_\perp^2 + m_h^2}
         z(1-z)\right), \nonumber \\*
P_{h-g+} &=& T_f \left((1 - z)^2 + \frac{m_h^2}{p_\perp^2 + m_h^2}
         z(1-z)\right). 
\label{226}
\ear
The splitting function is then:

\be
\Delta P_{hg} = P_{h+g+} - P_{h-g+} = \frac{1}{2}(z^2 - (1-z)^2),
\label{227}
\ee
which is protected against mass effects. That means 
that a decay of a gluon into a massive quark pair does not
produce any extra breaking of helicity. We can also get the 
unpolarized splitting function:

\be
P_{hg} = P_{h+g+} + P_{h-g+} = \frac{1}{2}(z^2 + (1-z)^2) + 
      \frac{m_h^2}{p_\perp^2 + m_h^2}z(1-z),
\label{228}
\ee
which agrees with the expression calculated in \cite{mrrs96}.

Our next step is to calculate the $\Delta P_{hh}$ splitting function:
In this case,

\be
\Delta P_{hh} =  C_F\frac{z(1-z)}{2} \frac{A}{p_\perp^2 + m_h^2(1-z)^2},
\label{229}
\ee
with

\be
A_{\pm +} = Tr[(\sla k_h + m_h)\frac{1 \pm \gamma_5}{2}
\gamma_\nu (\sla k_h^{\prime} + m_h) \frac{1 \pm \gamma_5}{2} \gamma_\mu]
\epsilon^* \cdot\epsilon.
\label{230}
\ee
The scalar product of the gluon polarization vectors is
$\delta^{ij} - k_g^i k_g^j/|\vec k_g|^2$, with $i, j =
1,2,3.$, and $k_g$ the gluon four momentum vector in the 
IMF.
In the calculation, we took the initial quark to be always
in a state of positive helicity.
For the helicity conserving part, $A_{++}$, we get exactly 
the same result of the massless (and also unpolarized) case.
For the helicity violating amplitude, we get $A_{-+} = -4m_h^2$.
It results in:

\be
P_{h-h+} = -2C_F z(1-z) \frac{m_h^2}{p_\perp^2 + m_h^2 (1-z)^2}
\label{231}
\ee
Hence,

\be
\Delta P_{hh} = C_F \left[\left(\frac{2}{1-z}\right)_+ -1 -z 
 + \frac{3}{2} \delta (1 - z) + 
 2 z(1-z) \frac{m_h^2}{p_\perp^2 + m_h^2 (1-z)^2} \right].
\label{231a}
\ee
A similar calculation shows that

\be
P_{g+h+} = C_F\left( \frac{1}{z} - z(1-z)
\frac{m_h^2}{p_\perp^2 + m_h^2 z^2}\right),
\label{232}
\ee
and

\be
P_{g-h+} = C_F \left( \frac{(1-z)^2}{z} - z(1-z)
\frac{m_h^2}{p_\perp^2 + m_h^2 z^2}\right).
\label{233}
\ee
Thus $\Delta P_{gh}$ is also protected from mass effects.
The same is not true for its unpolarized counterpart.

The last quantity to be calculated is the polarized partonic 
cross section, $\hat g_1^{(0)}$. It can be extracted from the
Born approximation to the antisymmetric part of the partonic
tensor. 

The general structure for the antisymmetric part of the 
partonic tensor is:

\be
\hat\omega_A^{\mu\nu} = - \frac{m_h}{2p.q}\varepsilon^
{\mu\nu\alpha\beta} q_\alpha[s_\beta \hat g_1 (z, Q^2) +
\left( s_\beta - \frac{s.q}{p.q}\right) \hat g_2 (z, Q^2)],
\ee
where $p$ is the parton four momentum vector, $s$ its 
spin vector and $q$ the photon four momentum vector.
If we define a general projector $P^{\mu\nu}
 = \varepsilon^{\mu\nu\alpha\beta} A_\alpha B_\beta$,
such that:

\be
P^{\mu\nu}\hat \omega_{\mu\nu}^A = -\frac{m_h}{2p.q}
(A.qB.s - A.sB.q) \hat g_1^{(0)},
\label{234}
\ee
then

\be
\hat g_1^{(0)}= -\frac{2p.q}{m_h} 
\frac{1}{A.qB.s - A.sB.q} P^{\mu\nu}\hat \omega_{\mu\nu}^A.
\label{235}
\ee
In the Born approximation, the antisymmetric part of 
the partonic tensor is

\be
\hat\omega_{\mu\nu}^A = m_h \varepsilon_{\mu\nu\alpha\beta} s^\alpha q^\beta
  \delta(2p.q + q^2).
\label{236}
\ee
The use of Eq. (\ref{236}) in Eq. (\ref{235}) then renders 
a coefficient function in LO which is unmodified by the presence of 
quark masses:

\be
\hat g_1^{(0)} = \delta (z - 1).
\label{237}
\ee

We see that in LO, the polarized splitting functions and the coefficient
function are protected against mass corrections, with the exception of
the $\Delta P_{hh}$ term. However, as shown in \cite{mrrs96}, any mass effect
appearing in the LO expressions, will be felt only in NLO.
This can be seen from the following. The general expression for the decay of a
parton in LO would be:

\be
dP(z) dz = \frac{\alpha(p_\perp^2)}{2\pi}\left\{
\frac{f(z)}{p_\perp^2 + m^2} + g(z)\frac{m^2}{(p_\perp^2 + m^2)^2}
\right\} dz dp_\perp^2.
\label{238}
\ee
As is well known, if we integrate Eq. (\ref{238}) $n$ times
in $p_\perp^2$
using $m^2 = 0$, with a strong ordering in $p_\perp^2$,
and sum over $n$, we obtain the
$[\alpha(\mu^2)/\alpha(Q^2)]^P$ of the renormalization 
group equations, 
where $P$ is the splitting function. When $m^2 \neq 0$, logarithms
in $\alpha$ are systematically lost, which is a typical feature
of NLO resummation.
%This is a typical feature of 
%NLO resummation, where the $n$ integrations give rise to a factor
%of the kind $[ln(\alpha_s(\mu^2)/\alpha_s(Q^2)]^{n-1} \alpha_s$. 
Besides this, the integrals of the mass terms in
Eq. (\ref{238}) are problematic, non analytical, 
and resummation is impossible.
In the same way, any mass modification of the NLO splitting 
functions, would appear only in a NNLO study.
With this in mind, we can then make a fully NLO evolution of 
a polarized heavy quark distribution using the same massless
expressions, the only exception being the LO $\Delta P_{hh}$
term.

The next step is to define the massive PGF cross section which 
interpolates the massless coefficient, $C_g$ in Eq. (\ref{13}), and the full 
massive coefficient, $H_g$ in Eq. (\ref{24}).
As noticed in Ref. \cite{aivazis94}, a part of $H_g$ has to 
be subtracted because
when evolving the heavy quark distribution, the $\Delta P_{hg}$
splitting function will also create heavy pairs. In the massless
evolution such a subtraction is already implicit in the absence 
of the $ln(Q^2)$ term. Using the slow evolution variable, 
$\xi = (1 + m_h^2/Q^2)x \equiv b x$, as defined by Brock \cite{brock80},
we write the heavy quark component as:

\be
\Delta h (x) \otimes \hat g_1^{(0)}(x) = \int_\xi^1 \frac{dz}{z}
\delta(z-1) \Delta h \left(\frac{\xi}{z}\right) = \Delta h(\xi).
\label{239}
\ee
Integrating the gluon splitting part of the LO DGLAP equation for 
$\Delta h$, we obtain:

\be
\Delta h(\xi) = \frac{1}{4\pi}\int_\xi^1 \frac{dz}{z} \int_{Q^2_{min}}^{Q^2}
dlnQ^{\prime 2} \alpha_s(Q^{\prime 2}) \Delta g (z, Q^{\prime 2}) 
\frac{\Delta P_{hg} (\xi/z)}{2}.
\label{240}
\ee
If we use $\alpha_s \Delta g$ as a scale invariant, then 
Eq. (\ref{240}) is reduced to:

\be
\Delta h(\xi) = \frac{\alpha_s (Q^2)}{4 \pi} ln\left(\frac{Q^2}{m_h^2}
\right) \int_{bx}^1 \frac{dz}{z} \Delta g(z, Q^2) 
\frac{\Delta P_{hg} (bx/z)}{2},
\label{241}
\ee
where we use $m_h^2$ for the minimum value of the scale.
According to the $\Delta P_{hg}$ splitting function, we should
start to evolve the heavy quark distribution at $m_h^2$. This
is explicit in Eqs. (\ref{222}) and (\ref{241}). And we actually
implement this scale in our calculations. 
%However, the resolution
%of charm is $4 m_c^2$ and the structure function, as a physical
%obeject, should respect this scale. In \cite{mrrs96}, an ad oc 
%factor $f = (1 - 4m_h^2/Q^2)\theta(1 - 4m_h^2/Q^2)$ was introduced
%into the heavy quark coefficient functions in order to force them
%to be nonzero only after $Q^2 > 4m_h^2$.
%In our case we will not introduce this factor, even because its
%functional form behaves as a free parameter, but we will calculate
%the structure functions only at scales where $Q^2 > 4m_h^2$.
However, to avoid any unconsistency with the scale necessary
for heavy quark to be resolved as a physical object (being probed by the 
photon), we will always calculate $g_{1p}^h$ at scales where
$Q^2 > 4m_h^2$. We then incorporate into the heavy quark 
coefficient function a threshold factor of the form 
$\theta (1 - 4m_h^2/Q^2)$. That means that below $4m_h^2$,
$g_{1p}^h (x, Q^2)$ is given by PGF only. As the 
choice of the threshold factor is not unique, the calculation of the heavy 
quark structure functions is, in a sense, ill defined. We return to 
this point in the next section.

Finally, we write down the convolution formula for the heavy 
quark component of the polarized structure function:

\bear
g_{1p}^h (x, Q^2) &=& e_h^2\int_{bx}^1 \frac{dz}{z} C_q (z, Q^2) 
  \Delta h\left(\frac{bx}{z}, Q^2\right) \nonumber \\*
  &+&
  e_h^2\frac{\alpha_s(Q^2)}{4\pi} \int_{ax}^1 \frac{dz}{z} H_g^{(1)}
  (z, m_h^2, Q^2) \Delta g(\frac{x}{z}, Q^2) \nonumber \\* 
  &-& e_h^2 \frac{\alpha_s (Q^2)}{4 \pi} ln\left(\frac{Q^2}{m_h^2}
\right) \int_{bx}^1 \frac{dz}{z} \frac{\Delta P_{hg} (z)}{2}
\Delta g\left(\frac{bx}{z}, Q^2 \right).
\label{242}
\ear
When we take $Q^2 >> m_h^2$ the last two  
lines of Eq. (\ref{242})
are reduced to the expression for massless quark production through PGF, and
we get Eq. (\ref{25}) corresponding to the massless case.
We notice that this is an exact result.

In light of the resulting Eq. (\ref{242}), it is instructive to analize the 
shortcomings of both MRRS and ACOT approaches for the calculation
of the heavy quark structure functions. The most evident 
and serious problem with MRRS is the fact that one does
not recover the massless, $\overline{MS}$ calculated,
unpolarized  gluon coefficient function when 
the limit $Q^2 >> m_h^2$ is taken \cite{mrrs96}. Hence, it can 
be hardly claimed that the defined heavy quark distributions
are $\overline{MS}$ distributions. In the same line,
$\overline{MS}$-like schemes are mass independent, meaning that  
evolution can not possibly depend on the mass of the 
quark, a requirement not fulfilled by the MRRS prescription.
However, because of the mass independence of both the polarized quark 
coefficient function and the polarized splitting functions in LO 
(the exception being $\Delta P_{hh}$ but it has no numerical 
relevance - see next section), Eq. (\ref{242}) is really the 
ACOT version for the calculation of the polarized heavy quark structure 
function. Hence, opposite to the unpolarized case, the polarized 
gluon coefficient function has the correct $\overline{MS}$ form 
in the limit $Q^2 >> m_h^2$. In a sense, it can be said that 
the ACOT formulation is preferable over the MRRS formulation for
$g_{1p} (x, Q^2)$. In any case, the ACOT formulation
also has formal drawbacks. Specifically,
the gluonic contribution to the structure function, which 
is of order $\alpha_s$, enters together with the
$\alpha_s^0$ quark contribution only, when the $\alpha_s$ quark 
corrections should also be considered. However, as stressed 
in the ACOT work \cite{aivazis94}, the $\alpha_s$ quark corrections
are numerically insignificant when compared to the gluon 
contributions. In the particular case of the present work, the 
$\alpha_s$ quark corrections were estimated, and their contributions are 
negligible. Finally, both schemes have ambiguities related to the  
choice of the factorization scale $\mu^2$, when physical 
observables, like structure functions, should not depend 
on the chosen value for $\mu^2$. Fortunately, in practical calculations
this dependence is not a real concern \cite{olness97}. 

\section{Results}

We are now ready to calculate $g_{1p}^h (x, Q^2)$. Although we will
restrict ourselves to charm, the extension to other heavy quarks 
is immediate. We will use the NLO GRSV \cite{grsv96} 
``standard scenario" parametrizations
for the gluon and quark singlet distributions as the input 
in our calculation. We made this choice 
because of the low scale where this parametrization is supposed to
be valid, which allows us to evolve through charm threshold. 
The strategy used for evolution was the following.
We first evolved the gluon and the singlet quark distributions from
0.34 $GeV^2$ to $m_c^2$, with 3 active flavours. From $m_c^2$ to 
$Q^2$, the charm quark distribution is obtained from the evolution of 
the singlet quark distribution with 4 flavours,
using as input distributions $\Delta\Sigma (z, m_c^2) = 0$ and
the $\Delta g(z, m_c^2)$ obtained from the evolution from 0.34 $GeV^2$.
The charm distribution at $Q^2$ is then $\Delta c(z, Q^2) = \Delta\Sigma
(z, Q^2)/4$. The gluon distribution at $Q^2$ is obtained from the 
singlet evolution of the gluon distribution with four 
active flavours, using as input distributions
$\Delta\Sigma (z, m_c^2)$ and $\Delta g (z, m_c^2)$, which were
obtained before when evolving from 0.34 $GeV^2$. Once we obtain the gluon
and charm distributions at $Q^2$, we can use the various convolution
formulas, Eq. (\ref{21}) for the pure PGF evolution via FOPT, 
Eq. (\ref{242}) for the improved MRRS method, and Eq. (\ref{25}) for 
massless evolution. 

We use $m_c = 1.5\; GeV$ for the charm mass, and 
$\Lambda_{QCD}^{(4)} = 0.2\; GeV$ in order to agree with GRSV.
The coupling constant at any scale, from $\mu^2$ to $Q^2$, is
given by the solution of the NLO equation for $\alpha_s$:

\be
\frac{\alpha_s(Q^2)}{4\pi} = \frac{1}{\beta_0 ln(Q^2/\Lambda_{QCD}^2)}
- \frac{\beta_1}{\beta_0^3} \frac{ln(ln(Q^2/\Lambda_{QCD}^2))}
{(ln(Q^2/\Lambda_{QCD}^2))^2}.
\label{31}
\ee
We notice that the use of this equation at low values of $Q^2$ such 
as 0.34 $GeV^2$ can lead to difficulties. In this case, the use of the 
transcendental equation for $\alpha_s$ from which Eq. (\ref{31}) is 
derived, would be more appropriate. However, to be consistent with 
the way the parametrizations were built, Eq. (\ref{31}) have to be used.

In Fig. \ref{fig1} we present the axial singlet charm distribution of the
proton at 10 $GeV^2$. It is obtained as explained before, evolving
the singlet distribution from $m_c^2$. This result shown is independent of
the mass modified $\Delta P_{qq}$ splitting function. It means that 
even if $\Delta P_{qq}$ is the only term which receives any mass modification,
it is also protected against mass terms for practical purposes. 
It is possible to fit the charm distribution as:

\begin{figure}[htb]
\graphics{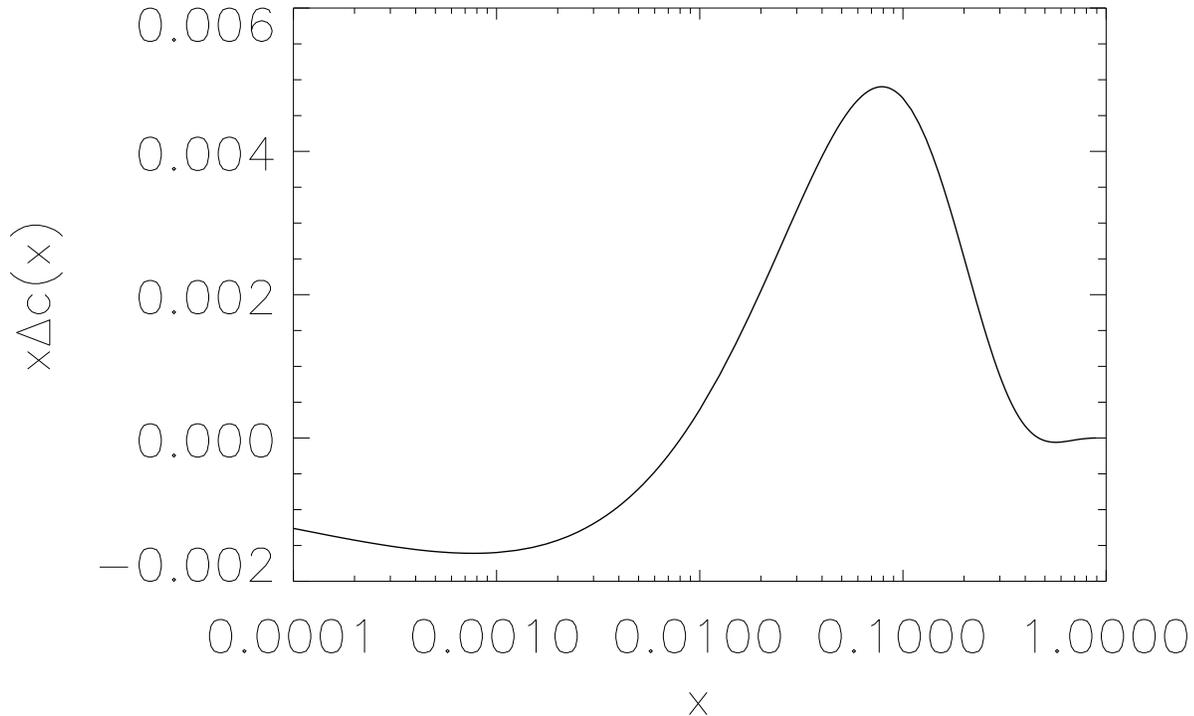}{20}{10}{-2}{-10}{0.9}
\caption{The $x$ dependence of the charm singlet axial charge at 
10 $GeV^2$.}
\label{fig1}
\end{figure}

\bear
x\Delta c (x, 10\; GeV^2) &=& x^{0.65} (-0.055 (1 - x)^5 + 0.206 (1 -x)^6)
 \nonumber \\*
 &+& x^{0.29}(-0.286 (1 - x)^7 + 0.261 (1 - x)^8).
\label{32}
\ear
The particular form for Eq. (\ref{32}) was inspired by Brodsky et al. 
\cite{brodsky95} considerations on the behaviour of the polarized distributions.
The integral over $x$ of Eq. (\ref{32}) is

\be
\Delta c (10\; GeV^2) \sim 2.6 \times 10^{-5},
\label{33}
\ee
which is completely irrelevant. However, we notice that 
charm is allowed in a large portion of the existing data
for $g_{1p}$, meaning that the study of the $x$ dependence 
is relevant, mainly in the small $x$ region.
 
\begin{figure}[htb]
\graphics{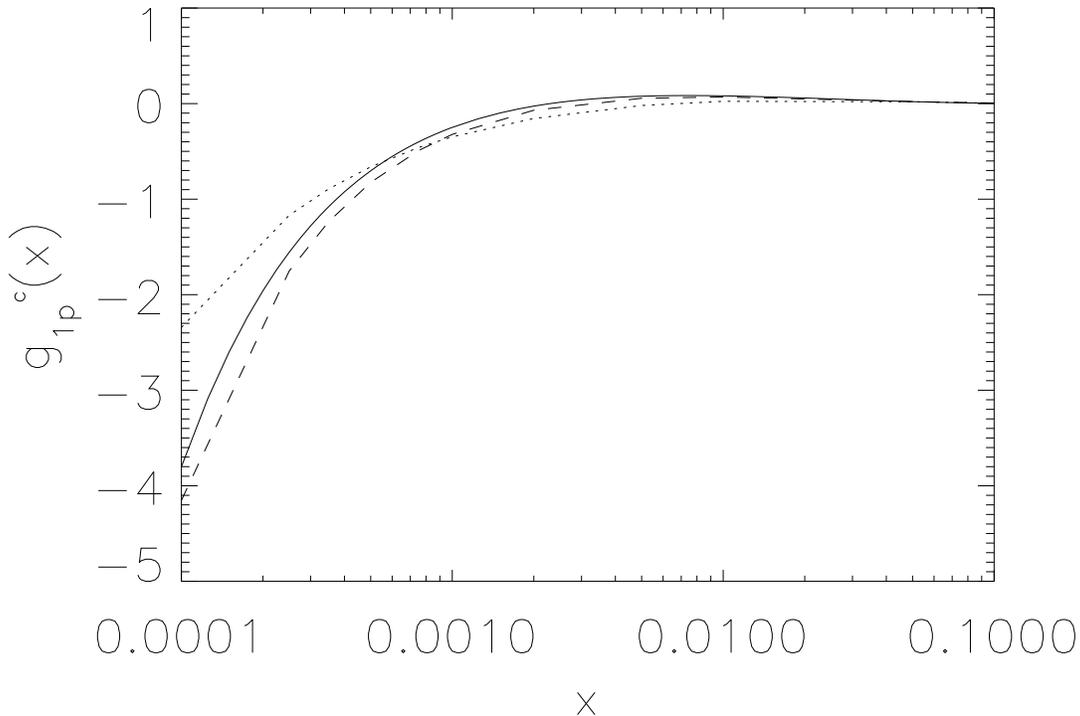}{20}{10}{-2}{-10}{0.9}
\caption{The charm component of $g_{1p} (x, Q^2)$ at 10 $GeV^2$.
The dotted line is the result from FOPT. The dashed line is the
result from massless evolution and the continuous line represents 
Eq. (\protect\ref{242}).}
\label{fig2}
\end{figure}

In Fig. \ref{fig2} the charm component
of $g_{1p}$ at 10 $GeV^2$ is presented. 
The dotted line is the result from FOPT, through 
Eq. (\ref{21}). In this case, $\mu^2 = 2.25\; GeV^2$ and $\alpha_s$
is calculated
using $n_f=4$. The dashed line is the result from massless evolution. The
continuous line is obtained with the help of Eq. (\ref{242}), which is the
scheme where low and high $Q^2$ are consistently incorporated. We see that
down to $x \sim 4\times 10^{-4}$, the three methods of calculating 
$g_{1p}^c (x, Q^2)$ give very close results. However, as $x$ gets smaller 
FOPT fails to follow the other two methods. We think this failure happens because 
at $x = 10^{-4}$, $\alpha_s (\mu^2) ln((1-x) Q^2/x m_c^2) \sim 3$, and
the NLO corrections calculated in \cite{buza96} would need to be incorporated.

\begin{figure}[htb]
\graphics{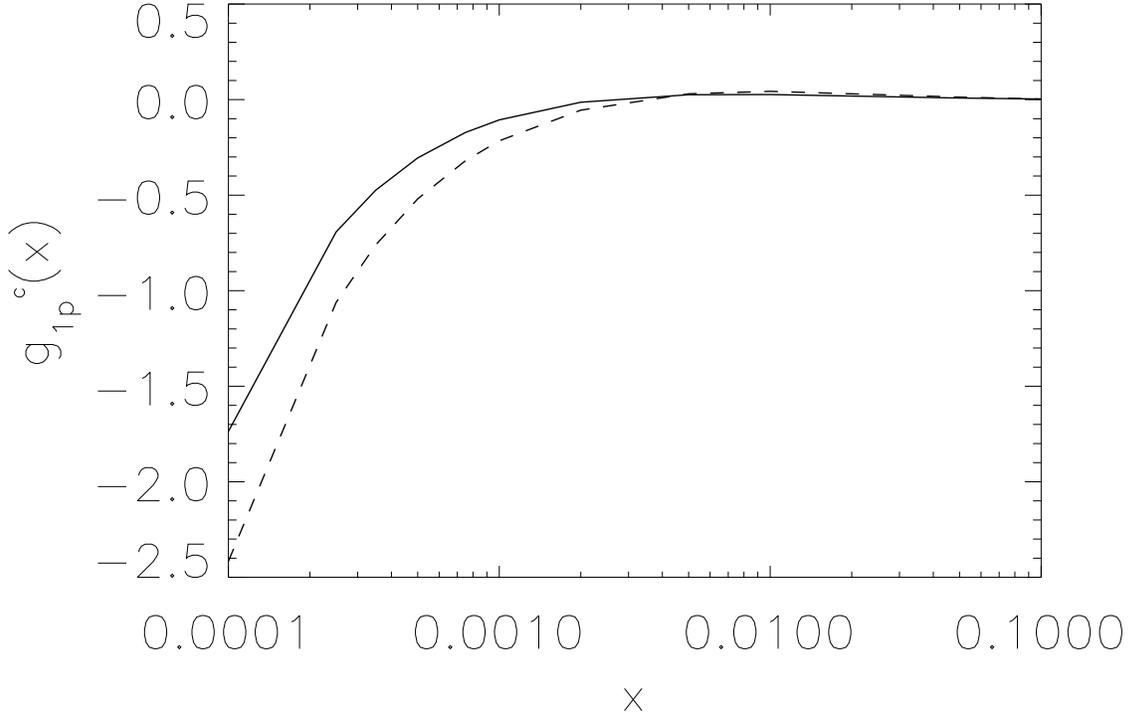}{20}{10}{-2}{-10}{0.9}
\caption{The continuous line is the charm component of Eq. 
(\protect\ref{242}) while the dashed line is the PGF term only.
Both curves are calculated at 10 $GeV^2$.}
\label{fig3}
\end{figure}

On the other hand, we plot in Fig. \ref{fig3} the $x$ dependence of the pure
PGF term, second line of Eq. (\ref{242}) and the pure charm quark distribution,
the first plus the third lines in Eq. (\ref{242}). The dashed line is the PGF
contribution and the continuous line the charm contribution. Comparing with
Fig. \ref{fig2}, we see that the PGF alone would reproduce the FOPT calculation,
which is a natural result as Eq. (\ref{21}) is pure PGF. But the charm contribution 
is almost as large as the PGF, hence the discrepancy between the 
dotted and continuous line of Fig. \ref{fig2}. That happens because, as stated
in the end of section 3, we choose not to add any ad hoc threshold factor
in the charm coefficient functions. If we had added, for instance, the 
factor $f = (1 - 4m_c^2/Q^2)\theta(1 - 4m_c^2/Q^2)$ as it is done in \cite{mrrs96}, 
the charm quark component of $g_{1p}^c (x, Q^2)$ shown in Fig. \ref{fig3} 
would be dramatically reduced.
In this case, the  line of Fig. \ref{fig2} would be very close to the 
FOPT result. However, the factor $f$ is arbitrary and other choices for it 
would produce different contributions from the charm evolution. In this sense,
the calculation of $g_{1p}^c (x, Q^2)$ at threshold is ambiguous. In our case,
we made the choice that if charm can be resolved, then it is given directly
by Eq. (\ref{242}). In other words, our threshold factor is the $\theta$ function
only.

\begin{figure}[htb]
\graphics{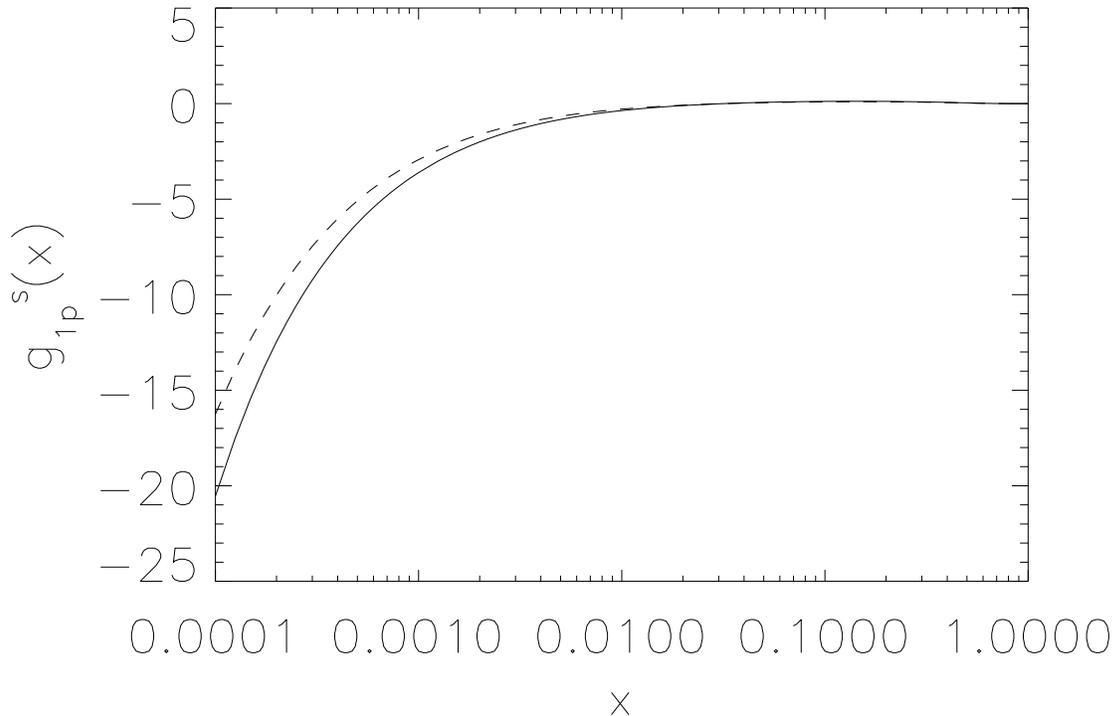}{20}{10}{-2}{-10}{0.9}
\caption{The singlet part of $g_{1p} (x, Q^2)$. The continuous
curve was evolved with 4 flavours, while the dashed curve
was evolved with 3 flavours. Both curves were calculated at
10 $GeV^2$.}
\label{fig4}
\end{figure}

We show in Fig. \ref{fig4} the result of evolving the singlet part of 
$g_{1p} (x, Q^2)$ with three and four flavours. The procedure used was
as the same used before. For three flavours, we evolve directly 
from 0.34 $GeV^2$ to 10 $GeV^2$, using $n_f=3$ and $\Lambda_{QCD}^{(3)} =
0.248\; GeV$. For the four flavours component, we evolve from 
0.34 $GeV^2$ to $m_c^2$ in the three flavour theory and from 
$m_c^2$ to 10 $GeV^2$ in the four flavour theory. Of course, the 
procedure is not completely consistent as the GRSV parametrization was
built using $n_f = 3$ only. However, it is expected that at $m_c^2$ the
parametrization is valid, meaning that the results shown here 
really point to the modifications
that a fourth flavour would induce to $g_{1p}^S (x, Q^2)$. Indeed, we see that
at $x=10^{-4}$, a difference of the order of 30\% is present.

\begin{figure}[htb]
\graphics{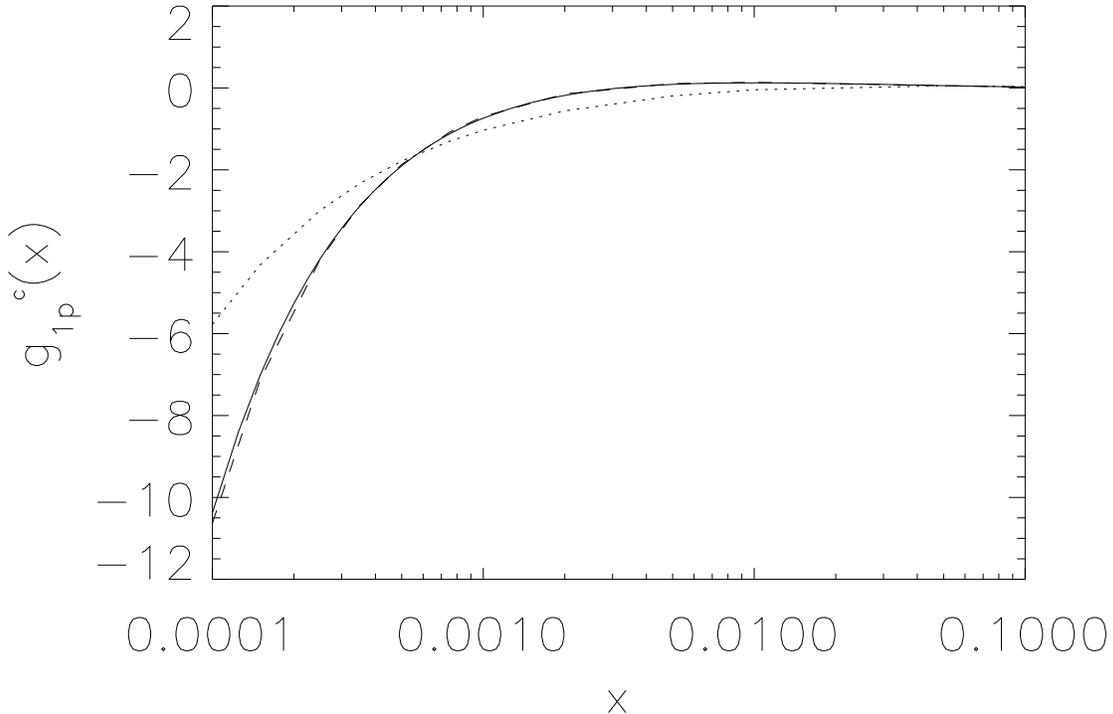}{20}{10}{-2}{-10}{0.9}
\caption{The same convention of Fig. 2, calculated now at 
100 $GeV^2$.}
\label{fig5}
\end{figure}

Finally, Fig. \ref{fig5} shows $g_{1p}^c (x, Q^2)$ at 100 $GeV^2$.
As expected, at this scale the improved equation for massive convolution
reproduces exactly the massless approach. FOPT, however, presents the
same sort of behaviour shown at 10 $GeV^2$. The reason for this
behaviour is the same as explained before.

\section{Final Remarks and Discussion}

We have computed the charm component of the polarized 
structure function of the proton, $g_{1p} (x, Q^2)$. We used
three different methods in these calculations and it is remarkable
that already at 10 $GeV^2$, massless evolution 
is a very good approximation. Of course, this conclusion is 
blurred by the umbiguous appearance of a possible threshold 
$f$ factor in the heavy quark coefficient functions. In any case,
it is seen that FOPT is close to massless evolution down 
to $x = 10^{-3}$, at least up to $Q^2 = 100\; GeV^2$. This conclusion
complements an early observation \cite{ziljstra94} where it 
was argued that FOPT and DGLAP evolution were equally good.
In truth, we see that when $x$ gets smaller, 
the large logarithms associated with the scale are enhanced by
the $1/x$ factors, something which does not happen in the usual
evolution approach or in the improved method of Eq. (\ref{242})
because the scale logarithms are absent.
FOPT then breaks down. Moreover, we saw in Fig. \ref{fig4}
how important the inclusion of a heavy flavour can be for the 
singlet component of $g_{1p}$, even if the amount of $\Delta c$ in 
the proton is small. 
Finally, when calculating the mass corrections to the splitting
functions, we saw that they are protected against mass effects.
Thus, effectively, Eq. (\ref{242}) is the polarized version of the
ACOT scheme \cite{aivazis94} for heavy quarks.
\newline
\newline
\newline
I would like to thank to F. S. Navarra, W. Melnitchouk and 
A. W. Thomas for the many discussions on the subjects covered in this
work as well as for a careful reading of this manuscript. 
I also would like to thank to T. Weigl for the basic code for NLO
evolution. Furthermore, I profited from the access to the computer
system of the Centre for the Subatomic Structure of Matter in Adelaide,
Australia. Finally, I would like to thank the Instituto de F\'{\i}sica
Te\'orica (IFT) for their support and
hospitality, and where part of this work was completed.
This work was supported by FAPESP (Brazil).

\addcontentsline{toc}{chapter}{\protect\numberline{}{References}}

\end{document}